\documentclass[]{tMOP2e}

\begin{document}




\title{Collisional Redistribution Laser Cooling of a High Pressure Atomic Gas}

\author{{Ulrich Vogl$^{a \ast \dagger}$\thanks{$^\ast$Corresponding author. Email: ulrich.vogl@nist.gov}\thanks{$^\dagger$Present address: National Institute of Standards and Technology and Joint Quantum Institute, Gaithersburg, MD, 20899, USA}\vspace{1pt}}, Anne Sa{\ss}$^{a}$, Simon Ha{\ss}elmann$^{a}$ and Martin Weitz$^{a}$\\\vspace{6pt} $^{a}${\em{Institut f\"ur Angewandte Physik, Wegelerstr. 8, 53115 Bonn, Germany}}\\\vspace{6pt}\received{v1.0} }

\maketitle

\begin{abstract}
We describe measurements demonstrating laser cooling of an atomic gas by means of collisional redistribution of radiation. The experiment uses rubidium atoms in the presence of several hundred bar of argon buffer gas pressure. Frequent collisions in the dense gas transiently shift a far red detuned optical field into resonance, while spontaneous emission occurs close to the unperturbed atomic transition frequency. Evidence for the cooling is obtained both via thermographic imaging and via thermographic deflection spectroscopy. The cooled gas has a density above 10$^{21}$ atoms/cm$^3$, yielding evidence for the laser cooling of a macroscopic ensemble of gas atoms. 
\end{abstract}

\bigskip

\section{Introduction}
Already in 1929 Pringsheim discussed an early concept to cool matter based on optical radiation \cite{Pringsheim}. Quite naturally, the general idea to use laser radiation for a cooling of matter is almost as old as the laser itself, and in the first years after the development of the laser in 1960 a few proposals have discussed ideas for possible realisations of laser cooling \cite{murao,dousmanis,Kushida}. The laser cooling scheme with the by far most profound impact is of course Doppler cooling of an atomic gas \cite{Hansch}, which utilizes momentum transfer from the laser photons to atoms in repeated absorption and subsequent spontaneous emission of photons. For atomic species with appropriate cooling transitions, a dilute gas can be cooled easily to the Microkelvin regime  \cite{nobelchu,nobelcohen,nobelphillips}.

Another class of laser cooling is Anti-Stokes cooling in multilevel systems, which can be applied for materials with strong Anti-Stokes transitions that are not accompanied by equally strong (heating) Stokes transitions or large quenching cross sections. Proof of principle experiments for this class of laser cooling have been carried out with a $CO_2$ gas \cite{djeu,bertolotti} and a dye-solution \cite{zander} respectively, but major advances have been accomplished in solid state materials as highly pure rare-earth doped glasses \cite{mungan,clark}, for which recently cooling down to 150\,K temperature could be realized \cite{Sheik-Bahae2}.

This paper discusses laser cooling by collisional redistribution of radiation, a technique that can provide cooling of a gas with density more than ten orders of magnitude greater than the typical values used in Doppler cooling experiments \cite{cooling}. Collisional redistribution is probably most widely known in the context of magnetooptical trapping of ultracold atoms, where this mechanism is a primary cause of trap loss processes \cite{adams}. In the long investigated field of room temperature interatomic collisions, redistribution of fluorescence is a natural consequence of line broadening effects from collisionally aided excitation \cite{yakovlenko}. In theoretical works, Berman and Stenholm in 1978 proposed laser cooling and heating based on the energy loss and gain respectively during collisionally aided excitation of atoms \cite{Berman_cool}. Experiments carried out with gases of moderate pressure have observed a heating for blue detuned radiation \cite{Berman_giacobino}, but the cooling regime was never reached.
In a recent work, we have demonstrated laser cooling of an atomic gas based on collisional redistribution, using rubidium atoms in argon buffer gas at a pressure of 230\,bar \cite{cooling}.
Frequent rubidium-buffer gas collisions in this high pressure gas environment shift a far red detuned laser beam transiently into resonance with a rubidium electronic transition, while spontaneous decay occurs near the unperturbed atomic transition. In such excitation cycles energy is extracted from the probe ensemble. In our proof of principle measurement with a thermally non-isolated sample, we demonstrated relative cooling of an atomic gas by 66\,K. The cooling power here reaches 87\,mW, corresponding to a cooling efficiency of 3.6\%. The aim of the present manuscript is to give a more detailed account of the experiment, and to discuss further experimental possibilities of the method.

\section{Principle of collisional redistribution cooling}

Our experiment involves rubidium atomic vapor in the presence of several hundred bar of buffer gas, where the collisionally broadened linewidth approaches the thermal energy $k_BT$ in frequency units. The general concept of the here described laser cooling approach is sketched in Figure 1, 
where the variation of the rubidium 5S ground state and the 5P excited state versus distance from an argon perturber atom are indicated in a binary collision quasimolecular picture.
When an argon atom approaches, the rubidium electronic D-line transition can be
 shifted into resonance with an otherwise far red detuned laser beam. This type of optical collision is sometimes also referred to as collisionally aided radiative excitation (CARE) \cite{yakovlenko, jablonski2, jablonski_prings,garbuny, Levin, Goela}. For the here relevant case of far red detuned radiation (we typically work with detunings of 10-20\,nm from resonance, i.e. far outside the usual rubidium thin vapor absorption profile), excitation is energetically only possible during a collisional process, and the excitation rate is hence dependent on the collisional rate. 
\begin{figure}
	\centering
		\includegraphics[width=0.5\textwidth]{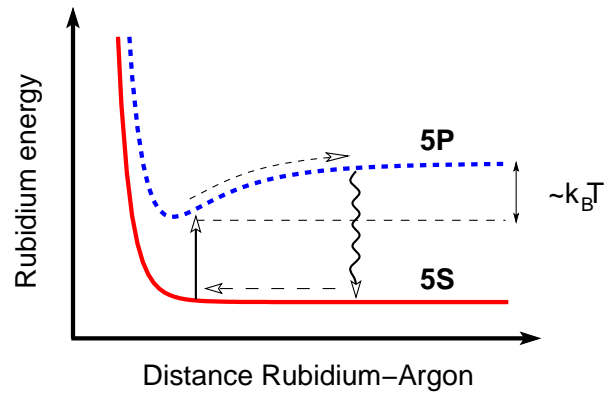}
	\caption{Cooling scheme. Indicated is the variation of atomic rubidium energy levels versus the distance to an argon perturber. Excitation with far red detuned light becomes possible during a collisional process.}
	\label{fig:setnatger}
\end{figure}
The rubidium excited state lifetime in the noble gas environment is in contrast largely unaffected by the 
frequent buffer gas collisions. 
After an excitation, spontaneous decay occurs most likely at larger distances between a rubidium atom and a perturber, i.e. close to the unpertubed atomic resonance frequency, so that the emitted photon is blue shifted relatively to the incident wavelength. This sponaneous decay completes the cooling cycle.
When the emitted photon can leave the sample, the atomic ensemble looses kinetic energy and will be cooled. The typical energy that can be removed from the reservoir in a cooling cycle is of order $k_BT\simeq h \cdot 10^{13}$\,Hz at T=500\,K, corresponding to a factor 20000 above the typical energy difference $\hbar k\cdot v$ removed by a fluorescence photon in usual Doppler cooling of dilute rubidium atomic gases.

\section{High pressure buffer gas spectroscopic setup}
 The experiment is performed in a high pressure stainless steel cell with volume of a few milliliters and sapphire optical windows. Two different cell designs are used. The first is a homemade design, where the sapphire windows are flanged with a nickel knive-edge seal to the cell body, and  the second one is a commercially available optical high pressure cell which uses Bridgman-sealing (Sitec-Sieber Engineering AG). The homemade design, which was used in most of the here described experimental measurements, allows for a quicker change of the optical windows, which makes it possible to vary the optical path length in the cell. More recently, we also have begun to use the commercial cell design, which in our experience in general is more longterm reliable at temperatures above 500\,K and pressures above 200\,bar. The pressure in the cell can be directly monitored via attached temperature stable steel-only manometers. 
After ultra-sound cleansing and a bake-out period the cells are filled with rubidium metal in natural isotope mixture (1\,g ampules) and argon buffer gas (see \cite{pra}). The cells are heated with standard high temperature heating wires and using a temperature  feedback control. The temperature of the measurement cell is typically chosen to be 10\,K higher than that of the rubidium reservoir part of the cell, to avoid rubidium condensation on the optical windows.
After carefully heating up the thermally isolated cell gradually over a period of typically one day, the equilibrium rubidium vapor pressure establishes accordingly to the set temperature \cite{nes}. 
The build-up of the vapor pressure can be monitored by optical absorption spectroscopy. We note that no satisfactory spectra were observed when e.g. the bake-out procedure used for the cleaning was not sufficient, or when the used buffer gas had too high levels of impurities (we use argon gas with a purity 6.0), which is understood to be able to cause quenching e.g. in the case of nitrogen admixtures.
 As primary laser source we use a continuous-wave titanium-sapphire laser system of 3\,W output power near the wavelength of the rubidium D-lines, whose output is directed into the pressure cell through the sapphire optical windows.

\section{Spectroscopic characterization of the rubidium-argon mixture}
Our cooling experiment typically operates with
a gas mixture of atomic rubidium vapor with number density in the order of 10$^{16}$cm$^{-3}$, corresponding to 1\,mbar vapor pressure for the used temperatures in the order of 500\,K, and argon buffer gas with a pressure in the range of 100-400\,bar (number density 1.5$\cdot$10$^{21}$cm$^{-3}$-6$\cdot$10$^{21}$cm$^{-3}$). The rubidium trace gas acts hereby as the optically active part, whose D-line transitions near 795\,nm and 780\,nm, respectively, are accessible with the titanium-sapphire laser. For characterization of the system by fluorescence spectroscopy, we used a confocal setup, as shown in Figure 2. This experimental scheme was applied to filter out reflections from e.g. the cell windows.
\begin{figure}
	\centering
		\includegraphics[width=0.5\textwidth]{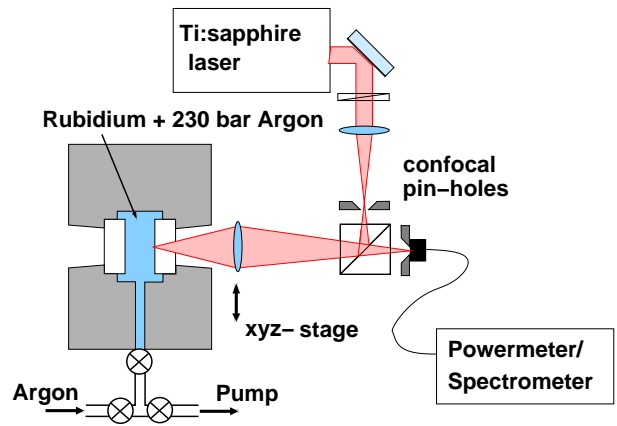}
	\caption{Experimental setup used in the case of the fluorescence spectra measurements, for which the light scattered by the atoms was detected in backward direction with a confocal geometry.}
	\label{fig:setnatger}
\end{figure}
Figure 3 shows a measurement of the collected fluorescence power using the confocal setup versus the wavelength of the incident laser beam. 
Clearly visible in the spectrum are the two D-line transitions near 780\,nm and 795\,nm. The resonances are strongly pressure broadened by the argon buffer gas, to typical linewidths of a few nanometers. 
For the shown case of the Rb-Ar pair, the lineshape is asymmetric, with an increased absorption amplitude on the red wing of the line. The large pressure broadening enables to efficiently excite the rubidium atoms at the large detuning values of order $k_BT/ h$ (corresponding to 23\,nm at T=500\,K) that were typically used in the cooling experiment. 

Essential for the collisional redistribution cooling process is a high elasticity of the rubidium-buffer gas collisions, and thus the absence of strong exothermic decay channels, i.e. quenching.
It is well known that alkali-rare gas collisions are remarkably elastic \cite{speller}, see \cite{schmorantzer} for measurements of the here investigated Rb-Ar pair at moderate buffer gas pressure values. To confirm the low quenching cross section in the here investigated high pressure regime, where also multiple particle collisions are relevant, we have measured the lifetime of the rubidium 5P excited state in our system. 
\begin{figure}
	\centering
		\includegraphics[width=0.5\textwidth]{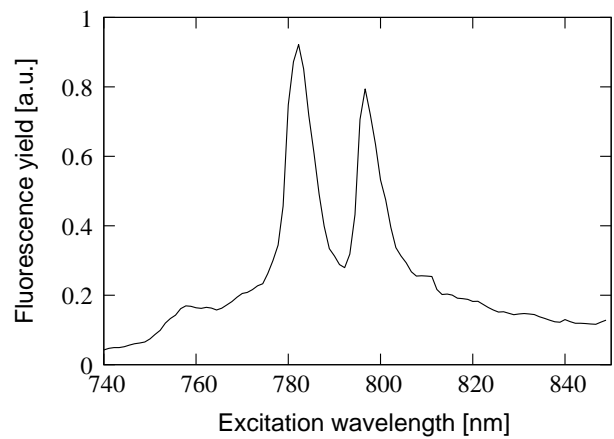}
	\caption{Measured fluorescence power for atomic rubidium vapor at 530\,K temperature with 230\,bar pressure of argon buffer gas, versus the excitation wavelength.}
	\label{fig:setnatger}
\end{figure}
At low buffer gas pressures ($\le$10\,bar) we observed fluorescence
decay times up to 1$\mu$s, which we attribute to repopulation from higher states due to energy pooling \cite{ekers}.  Besides this value being much above the known lifetime value of the rubidium 5P state, a  further indication for the occurance of energy pooling was the clearly visible blue fluorescence at those low pressure values (for example, radiation near 425\,nm, arising from fluorescence from the 6P state to the 5S ground state, was observed). At higher buffer gas pressures, the measured decay time of the fluorescence signal, if accounting for the experimental time constant of our detection system, approaches the known value 27\,ns for the natural lifetime of the 5P state. We conclude
that energy pooling to higher state excitation is suppressed
at sufficient buffer gas pressure. The measured $5P$ state lifetime is, within our experimental accuracy of 2\,ns, unchanged up to the used argon pressures near 230\,bar. At the applied temperature and buffer gas pressure, the collision rate is in the order of 10$^{11}$/s and the atoms experience more than 10$^4$ collisions within a natural lifetime. Our measurement accuracy here was mainly limited by the finite risetime of the acoustooptic modulator used in these measurements for beam switching. The measurement results clearly confirm the remarkable elasticity of collisions of excited state rubidium atoms with a rare gas, which is a prerequisite for the operation of collisional redistribution cooling.

\section{Redistribution of fluorescence}
We have measured the average optical frequency difference between incident and fluorescence light, which from energy conservation allows to estimate the expected cooling power extracted from the sample. For the corresponding measurement, we used the described confocal setup (see Figure 2), to selectively detect fluorescence from the volume of the cooling beam focal region. Typical spectrally resolved fluorescence spectra for exciting laser frequency of 357\,THz and 401\,THz, respectively, are shown in Figure 4. It is clearly visible that the detected fluorescence is spectrally redistributed to the center of the D-lines. The observed spectral width of the fluorescence radiation peaks (located at the two rubidium D-lines) is a few nanometer. The spectrum of the emitted fluorescence is largely independent of the used incident laser wavelength. The mean experienced energy shift per photon is thus roughly equivalent to the differential energy determined by the detuning of the incident laser light from resonance. For a more detailed investigation, Figure 5 shows the variation of the shift of the optical exciation frequency from the observed mean fluorescence frequency. In the investigated spectral range, the corresponding energy difference between excitation and emission can reach a few percent of the photon energy, which,
if we neglect heating processes from inelastic decay channels, gives an estimate for the expected energy transferred to the atomic ensemble per photon (the above discussed lifetime measurements give an upper boundary for the inelastic decay channels). For red detuned laser radiation, energy can be removed from the sample, so that the sample is cooled.

\begin{figure}
	\centering
		\includegraphics[width=0.6\textwidth]{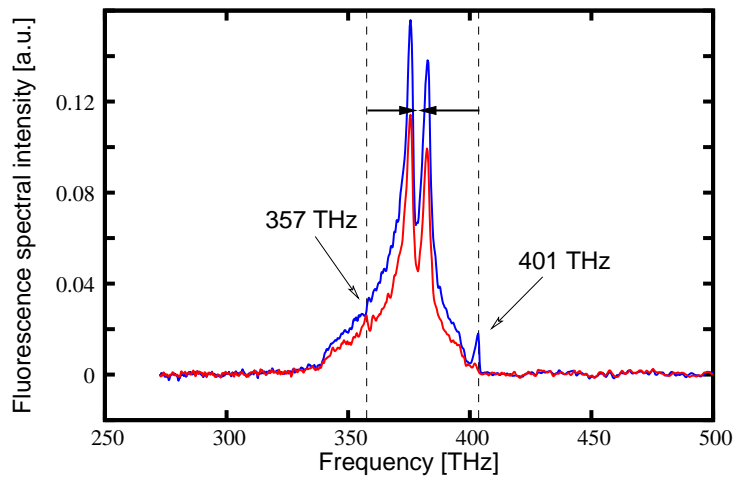}
	\caption{Spectrally resolved fluorescence signal of rubidium vapor at 530\,K with 230\,bar argon buffer gas. Shown are cases for one red detuned (357\,THz) and one blue detuned optical excitation frequency (401\,THz). The incident laser frequencies are indicated as vertical dashed lines. The shown arrows indicate the direction of the mean experienced frequency shift of the scattered photons by the redistribution process towards the D-lines center.}
	\label{fig:setnatger}
\end{figure}
\begin{figure}
	\centering
		\includegraphics[width=0.5\textwidth]{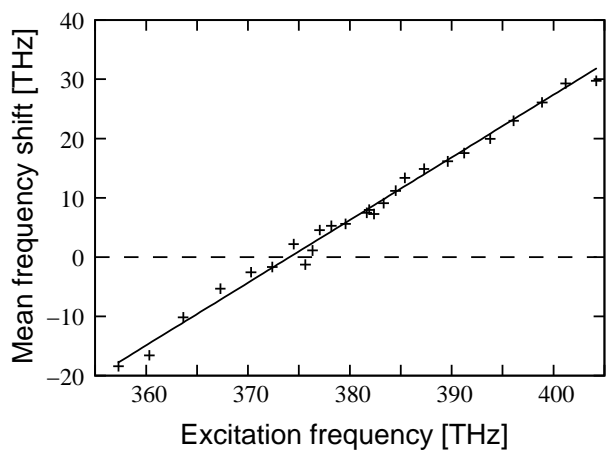}
	\caption{Frequency shift beween exciting and the measured mean frequency of the emitted fluorescence, versus the optical excitation frequency. The observed zero-crossing is slightly red to the D1-line, which is understood in terms of the pronounced red wing of the rubidium spectra in argon buffer gas.}
	\label{fig:setnatger}
\end{figure}

 The expected cooling efficiency per photon is
\begin{equation}
	f=\frac{\overline{\nu_{\text{fl}}}-\nu_{\text{laser}}}{\nu_{\text{laser}}},
\label{eq:nulaser}
\end{equation}
where $\nu_{\text{fl}}$ and $\nu_{\text{laser}}$ denote the centroid frequency of the fluorescence and the laser frequency respectively.
This allows
to estimate the cooling power that is withdrawn from the gas (or deposited for the case of heating as the result of a blue laser detuning). 
If $P_{\text{opt}}$ denotes the incident laser power and $a(\nu)$ the absorption probability in the gas for an incident laser frequency $\nu$, we arrive at an expected cooling power (that is, if we neglect inelastic processes) of
\begin{equation}
	P_{\text{cool}}=P_{\text{opt}}a({\nu})\frac{\overline{\nu_{\text{fl}}}-\nu_{\text{laser}}}{\nu_{\text{laser}}}.
\label{eq:pcool}
\end{equation}
The absorption probability depends on the frequency, and we have measured the dependence of the transmitted optical power on the exitation wavelength. On resonance the optical density of the cell reaches a maximum value of about 5, which decreases at the wings of the pressure broadened profile (following the spectral profile shown in Figure 3).
The deduced expected cooling power, according to Eq.\,2, versus the laser frequency is shown by the connected red vertical crosses in Fig.6.
 For a red detuning of the laser light of 10-20\,nm from resonance one can expect a significant cooling power in the order of 100 mW, which exceeds the cooling power reached in Doppler cooling experiments by many orders of magnitude.

\begin{figure}
	\centering
		\includegraphics[width=0.5\textwidth]{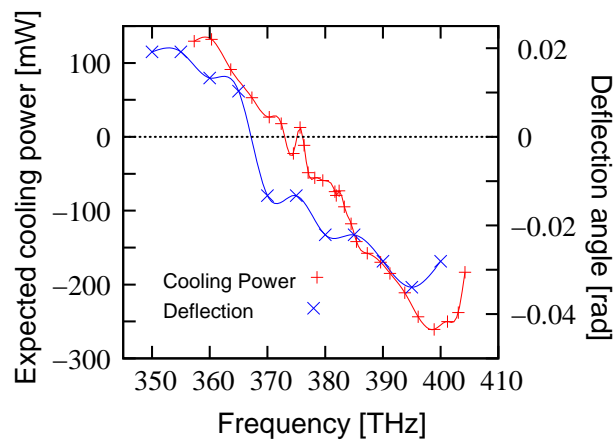}
	\caption{The connected red vertical crosses show the expected cooling power (left-hand axis) as a function of laser frequency, as derived from the average fluorescence frequency shift. The blue crosses (fitted with a spline function) show the angular deflection (right-hand axis) of a probe beam.}
	\label{fig:setnatger}
\end{figure}

\section{Thermographic detection}
The above shown data indicate that a cooling of the dense gas sample is possible by collisional redistribution, but do not yet give a proof of cooling, despite the described lifetime measurements. The main reason is that we a priori cannot exclude heating processes from e.g. other residual gases (which in the case of nonradiative deexcitation would heat the sample by an amount of $h\nu$ per photon, while the cooling efficiency per redistributed photon is $\textit{f}\cdot h\nu\simeq k_BT$, i.e. has an efficiency of a few percent). For a direct detection of the cooling, noncontact temperature measurements have been used. Initial attempts to directly detect cooling by using commercial thermocouples were unsuccessful, mainly because absorption of the fluorencence radiation scattered from the atoms resulted in a heating of the thermocouple material largely independent of the laser frequency. We conclude that the application of thermocouples near to the cooling region would require the use of highly reflecting optical surfaces. But for the here reported measurements, we have instead moved to a thermographic temperature detection scheme.
\begin{figure}
	\centering
		\includegraphics[width=0.6\textwidth]{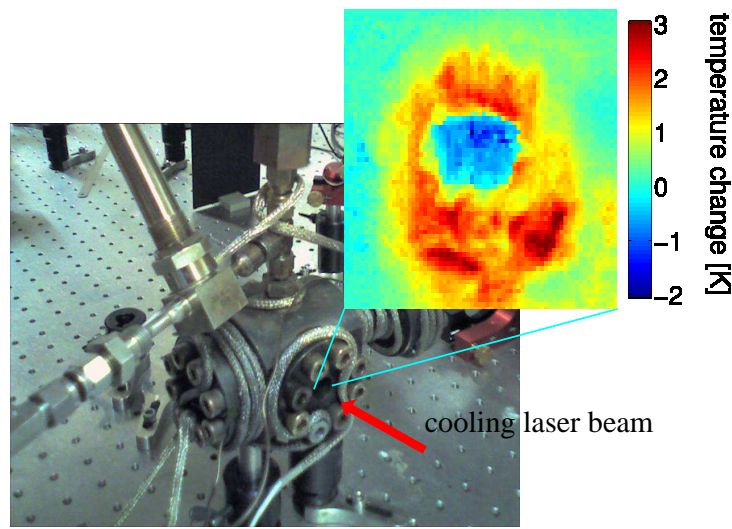}
	\caption{For a quantitative detection of the cooling, a thermographic camera was directed at the sapphire entrance window of the high-pressure buffer-gas cell. The image shown in the upper right gives the (average) colour-coded temperature change after a 30-s long pulse of the cooling beam. The blue region in the centre is the sapphire window, cooled from behind by the gas sample. The surrounding metal flange appears red and yellow, owing to heating from absorption of scattered fluorescence radiation. To avoid cooling laser radiation having a possible residual influence on the infrared-camera signal, we activated the camera only after the end of a cooling-laser pulse. The diameter of the cooling beam used in this measurement was 3\,mm. The main figure gives a photo of the used homemade high pressure cell. During operation, the cell is additionally coated with a mineral garment for thermal isolation.}
	\label{fig:setnatger}
\end{figure}

As became obvious, the used sapphire optical windows of our pressure cell have good thermal conductivity and are highly nonabsorbing at the cooling laser wavelength, which allows to use the windows as a temperature probe.
The first direct observation of cooling was obtained with an infrared camera (Flir systems, model  T200) sensitive to black-body radiation in the wavelength range of 7.5 - 13 $\mu$m, which we directed at the outer surface of a sapphire window of our pressure cell. In this mid-infrared spectral region, the sapphire window material is highly opaque and thus emits black-body radiation according to its temperature. Owing to thermal transport in the sapphire material (the thermal conductivity of sapphire is 20000 Wm$^{-1}$K$^{-1}$), a cooling of the dense gas will cause a temperature variation also at the outer surface of the cell window.  For this measurement, we determined the temperature change of the sapphire window following a 30-s optical-cooling period.
The diameter of the cooling beam used for the thermographic detection measurement was 3\,mm.
 Figure 7 shows the result recorded for a cooling-laser frequency of 365\,THz, corresponding to a red detuning of 16\,THz relative to the rubidium D-line centroid. The observed average temperature decrease is 0.31(0.03)\,K near the beam centre. The temperature increase of a metal flange of the cell is also clearly visible in the figure, and is attributed mainly to absorption of scattered fluorescence radiation. The observed temperature decrease at the cell window is in agreement with the result of a simple heat-transport model calculation.

\section{Thermal deflection spectroscopy}
To determine the temperature change of the gas inside the pressure cell, we use thermal deflection spectroscopy \cite{Boccara,spear,whinnery}. The cooling laser beam induces a local temperature change in the gas, and the main idea here is to determine the magnitude of this change from the refractive index profile experienced by a probe laser beam. In the used argon buffer gas, the Lorentz-Lorenz relation $\frac{1}{\rho}(\frac{n^2-1}{n^2+2})=\text{const.}$, where $\rho$ is the density of the gas, gives in good approximation an expected change of the refractive index $n$ of
\begin{equation}
	\frac{dn}{dT}\simeq-\frac{n-1}{T}.
\end{equation}
The transverse profile of the laser beam induces a nonuniform temperature profile that is accompanied by a corresponding profile of the refractive index, because temperature variations cause a change of the local gas density. A second nonresonant laser beam can probe this induced thermal lens.
\begin{figure}
	\centering
		\includegraphics[width=0.6\textwidth]{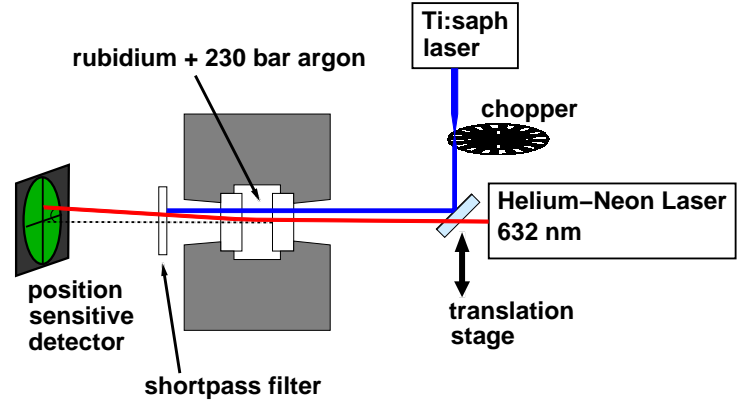}
	\caption{Experimental setup used in the thermal deflection spectroscopy measurements.}
	\label{fig:setnatger}
\end{figure}
The deflection of the probe beam along the beam path is given by
\begin{equation}
	\Phi=\frac{n-1}{T}\int^{L}_{0}\frac{dT}{dr}dz.
\end{equation}
The cooling beam has a $TEM_{00}$ profile with beam radius $w$ and experiences an absorption coefficient $\alpha$. Its intensity distribution is
\begin{equation}
	I(r,z)=\exp\left(-\alpha(\nu)z-\frac{2 r^2}{w^2}\right).
\end{equation}
For a heat transport that occurs mainly in the radial direction (we assume a long cell and that the beam focal radius is much smaller than the absorption length), and assuming that the cooling beam removes heat with a rate following its intensity distribution, the temperature distribution can be determined from the beam deflection using:
\begin{equation}
	\Delta T(r,z)=\frac{T}{n-1}\frac{\alpha e^{(-\alpha z)}}{1-e^{(-\alpha z)}}\int^{\infty}_{r}\Phi(r')dr',
\end{equation}
where r denotes the transverse offset between cooling and probe laser beams. For the refractive index of the Rb-Ar mixture at the 632\,nm probe wavelength, which is non-resonant with the rubidium D-lines, a value n$\simeq$1.028(4) for argon is used \cite{born}, where the quoted uncertainty is basically determined by the accuracy of the pressure measurement.

For comparison with the experimental results, the data is fitted with the expected temperature profile derived from a heat transfer model \cite{whinnery}. After a cooling time $t$, the expected temperature profile is
\begin{equation}
	\Delta T=\frac{P_{\text{cool}}}{4\pi\kappa}e^{-\alpha z}\left[\text{Ei}\left(\frac{2r^2}{w^2}\right)-\text{Ei}\left(\frac{2r^2}{w^2-8Dt}\right)\right],
\label{eq:four}
\end{equation}
where $\kappa$ is the specific heat capacity, $w$ the beam radius (the same for both cooling and probe beam),  $D$ the thermal diffusivity, and $P_{\textit{cool}}$ is the cooling power.
The Ei-function is a consequence of the heat diffusion with a Gaussian shaped cooling laser beam. For the temperature gradient one obtains the expression
\begin{equation}
	\frac{dT}{dr}=-\frac{P_{\text{cool}}}{2\pi\kappa}\frac{e^{-\alpha z}}{r}\left[\exp\left(\frac{2r^2}{w^2}\right)-\exp\left(\frac{2r^2}{w^2-8Dt}\right)\right].
\label{eq:four2}
\end{equation}
In our experimental setup, a (non-resonant) helium-neon laser probe beam is sent collinearly, and slightly displaced from the cooling beam, through the high-pressure cell, as indicated in Figure 8. Radial temperature variations induced by the cooling beam cause a gradient in the refractive index of the gas, and result in a prismatic deflection of the probe beam. The diameter of both the cooling beam and that of the detection beam used in the measurements reported here was 1\,mm each.
The blue crosses in Figure 6 show the measured deflection angle as a function of cooling-laser frequency. For red detuning, the probe beam is deflected towards the cooling beam, which indicates cooling, whereas for blue detuning heating is observed. It is also apparent that the cooling regime is only reached when the laser frequency is detuned approximately 5\,THz lower than the value at which it just equals the average fluorescence frequency; that is, where the expected cooling power given by Eq.\,\ref{eq:pcool} reaches zero. This is attributed to residual heating, which could be due to a small quenching cross section of excited-state rubidium atoms in the ultrahigh-pressure buffer-gas environment. Our data could be explained by an average decrease in the 27-ns natural lifetime of the rubidium 5P state due to quenching in inelastic collisions by roughly 1\,ns at an argon buffer-gas pressure of 230\,bar, which would be consistent with the directly measured value of the lifetime rubidium $5P$ state. This limit means that the ratio of elastic to inelastic collisions of excited-state rubidium atoms with argon rare-gas atoms here exceeds 20000:1.
\begin{figure}
	\centering
		\includegraphics[width=0.6\textwidth]{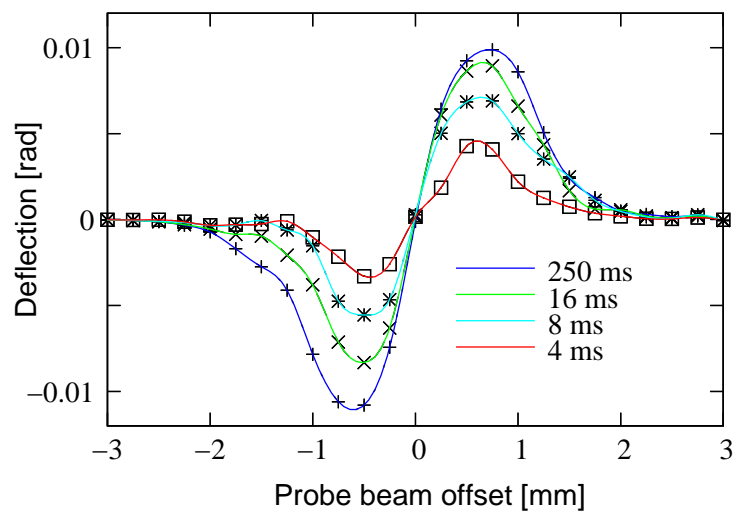}
	\caption{Deflection of the probe beam in the thermal deflection setup versus the probe beam position for different cooling times. The deflection is a direct measure for the experienced temperature gradient.}
	\label{fig:setnatger}
\end{figure}

To acquire the spatial temperature profile induced by the cooling beam, we scanned the lateral offset between the cooling and the probe beam. 
Typical obtained deflection traces are shown in Figure 9 for a cooling-laser frequency of 365 THz and a relative absorption of 90\% in the 1-cm-long gas cell for different exposure times.
The measurements suggest that basically a steady state value of the cooling is reached at the largest shown exposure time of 250\,ms, and the corresponding deflection data for this cooling time is again shown by the dots in Figure 10, along with a fit by a theoretical heat-transfer model (black line), see Eq.\,7. The cooling power and the thermal diffusivity here were left as free fit parameters. We derived the lateral temperature profile near the cell entrance from a numerical integration of the data over the transverse profile. The result is shown by the solid blue line in Figure\,10. The obtained temperature drop in the centre of the cooling beam is 66(13) K, starting from an ambient cell temperature of 620\,K. This value is much larger than the above-mentioned result for the temperature drop detected at the outer surface of the cell window. This is easily understood as resulting from the thermal conductivity of the argon gas being much less than that of the sapphire window material, meaning that the cooling within the gas volume near the focal region of the beam can proceed further. The dashed blue line in Figure 10 is the direct result of the model (Eq. 8), which yields a fully symmetric curve.
A cooling power of 87(20) mW can be estimated from the model, corresponding to a cooling efficiency of 3.6(0.8)\%. These values are in reasonable agreement with the expected cooling power derived from the differential frequency shift of the fluorescence photons (see Figure 5 and the corresponding discussion). We note that the cooling power is four to five orders of magnitude greater than that achieved in Doppler-cooling experiments, and is comparable to results reported for the laser cooling of solids.
\begin{figure}
	\centering
		\includegraphics[width=0.6\textwidth]{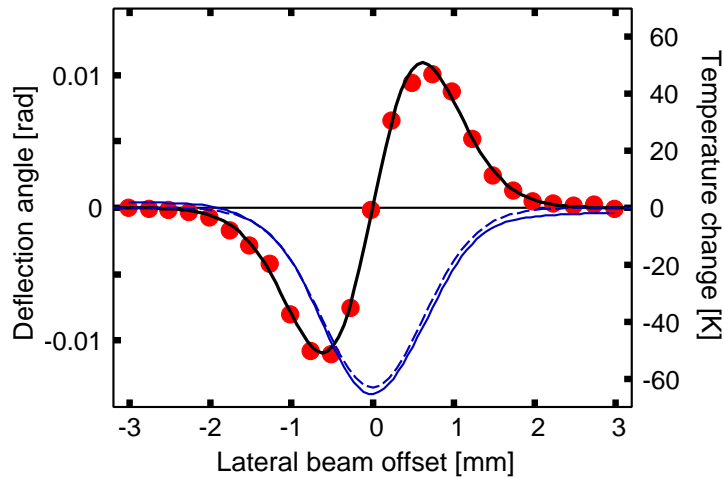}
	\caption{The dots show the deflection signal for 250\,ms cooling laser beam exposition. The temperature profile in the gas is obtained by integrating the temperature gradient, which can be deduced from the deflection signal. The solid blue line shows the corresponding temperature profile near the cell entrance. The determined temperature drop in the beam centre is 66(13) K. The quoted error bars are determined mainly by uncertainties in the optical-beam diameter, the beam geometry, the measured absorption coefficient for the cooling radiation and the buffer-gas pressure measurement. The dashed blue line follows a heat-transport model.}
	\label{fig:setnatger}
\end{figure}

The model we used to simulate heat transfer in the gas cell volume assumes that the cooling rate is proportional to the optical intensity of the Gaussian laser beam; that is, we neglect the influence of radiation transport, which assumes that $\omega \ll l_{abs}$, where $\omega$ denotes the cooling-beam radius and $l_{abs}=1/\alpha$ the absorption length. In general, we expect the influence of radiation transport to be weaker than in the area of ultracold atomic gases, as the influence of light forces in the buffer-gas sample is much smaller. 
The absorption length is related to the rubidium density $n_{Rb}$ by the relation $l_{abs}=1/\sigma n_{Rb}$, where we can estimate the resonant absorption cross section of the pressure broadened rubidium resonance by $\sigma=\frac{\lambda^2}{2\pi}\frac{\Gamma_{nat}}{\Gamma_{real}}$, with a ratio of the natural and the pressure broadened linewidth of $\frac{\Gamma_{nat}}{\Gamma_{real}}\simeq$10$^{-6}$ for our typical experimental parameters. Correspondingly, the absorption cross section is six orders of magnitude below its (resonant) value in the dilute vapor regime of $\frac{\lambda^2}{2\pi}$ . The rubidium vapor densities of typically 10$^{16}$\,cm$^{-3}$ used in the present experiment thus do not lead to a larger reabsorption, per given length scale, than a gas with density of 10$^{10}$\,cm$^{-3}$ in the absence of a buffer gas, as are the typical densities of a magnetooptic trap used in the area of ultracold, dilute atomic gases \cite{adams}.

\section{Redistribution cooling in the presence of fine structure changing collisions and multiparticle collisions}
We have restricted the discussion of redistribution cooling so far to a two-level model with one ground state and one excited state, and moreover assumed only binary collisions with the buffer gas atoms. In reality, the rubidium 5S-5P transition is split up by the upper state fine structure into the D1- and D2-lines at 377\,THz and 384\,THz respectively. 
Further, in the used high density buffer gas regime the binary collision assumption no longer is fully valid. The typical collision impact parameter (or optical collision radius) for alkali-noble gas pairs is of the order 0.5-1\,nm, see \cite{weisskopf}. The mean interatomic distance at a typical buffer gas density of $3\cdot 10^{21}$cm$^{-3}$ is only 7\,$\AA$, i.e. it almost reaches the order of magnitude of the size of the buffer gas atoms, and a description of the system in the form of alkali-exciplexes becomes relevant (relevant related investigations are currently made in the field of alkali-lasers (e.g. \cite{Readle}).

A scheme of the ground and the excited state potential curves of the rubidium-argom system including the 5P fine structure is shown in Figure 11 in a binary collisional picture. The corresponding potential curves have been calculated using the obtained values in \cite{pascale}. For the used detuning of a few nm red of the D1-line (that is, detuned to a much lower frequency), initially the 5P$_{1/2}$ state is excited by collisionally aided excitation. The frequent collisions (the collisional rate is 10$^{11}$/s) then cause efficient redistribution between the 5P$_{1/2}$ and 5P$_{3/2}$  potential curves within the nanoseconds long excited state lifetime. The fluorescence spectra of Figure 4 show the double-peaked resonance structure of both rubidium D-lines, as is expected from the redistribution between the upper state fine structure manifold. The redistribution can be understood from the typical exchanged energy per collision being of the order of $k_BT$, which for the ambient temperatures of 500\,K is larger than the energy gap between the fine structure levels. Following a large number of collisions, we expect that the two fine structure levels will be populated according to the detailed balance condition. We note that recent experimental work in the Rb-He system has shown that multiple particle collisions can lead to a mixing rate between the upper state fine structure levels that is enhanced over that expected in a binary collisional picture \cite{sell}. As the 5P$_{3/2}$ state is higher energetic than the 5P$_{1/2}$ state, the energy loss of the sample after the emission of a photon from this state is even larger than in the case of the 5P$_{1/2}$ state. One thus may expect that a fine structure splitting and multiple collisions could help in the cooling process, but we are aware that the influence of these effects needs to be studied in more detail in the future.
\begin{figure}
	\centering
		\includegraphics[width=0.7\textwidth]{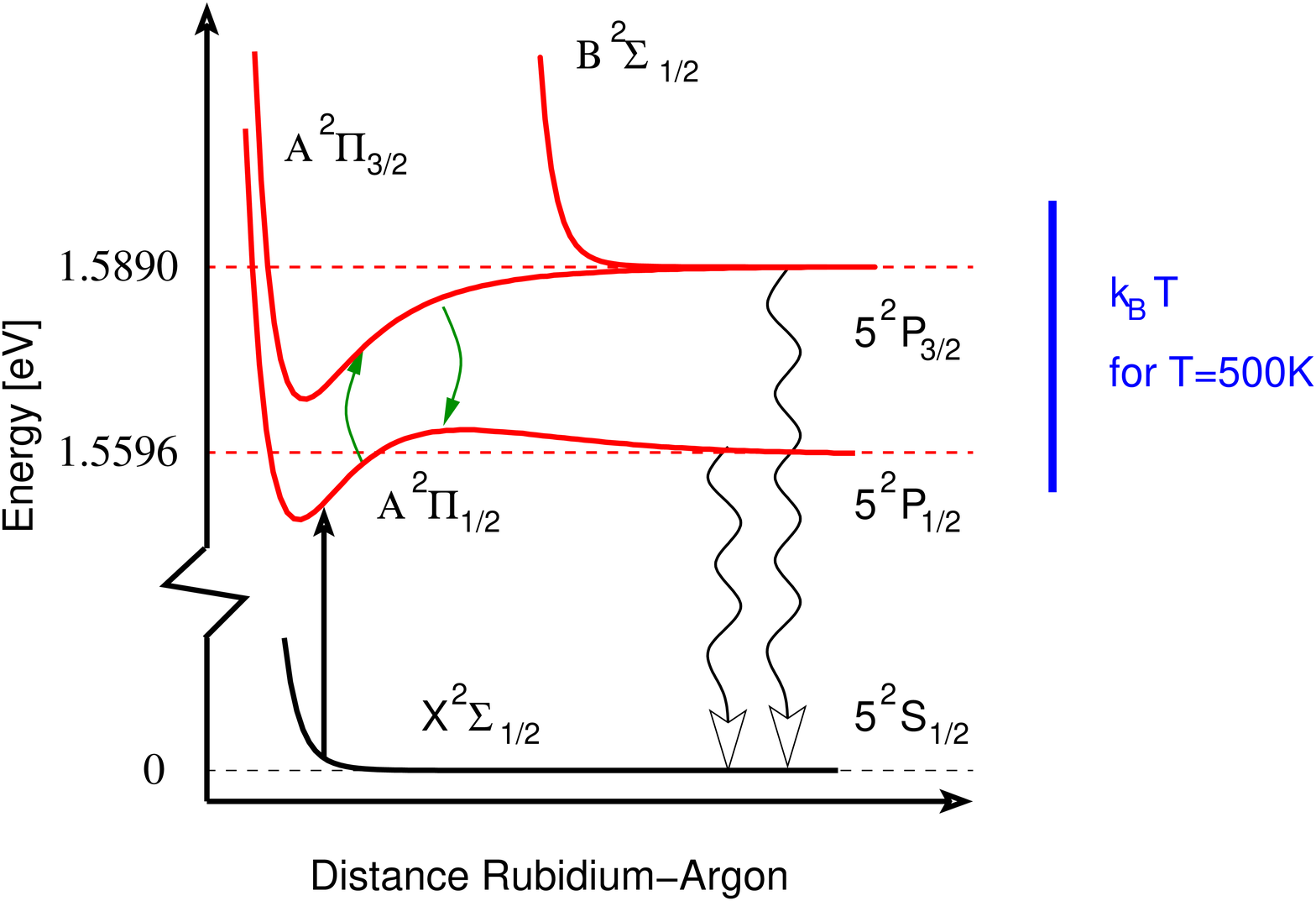}
	\caption{Ground and excited state potential curves for the Rb-Ar system, as calculated using parameters given in \cite{pascale}. Far red detuned radiation can be absorbed when an argon pertuber approaches the rubidium atom, leading to an excitation of the collisionally modified 5P$_{1/2}$ state. The collisions also cause efficient redistribution between the upper state fine structure manifold, as is indicated by the green arrows, resulting in double-peaked fluorescence spectra, as shown in Figure 4. The vertical blue line indicates the size of the thermal energy $k_BT$ at a typical cell temperature.}
	\label{fig:setnatger}
\end{figure}

The primary reason for choosing a pressure range of a few 100\,bar for the present experiments was to make the collisional broadening comparable to the thermal energy $k_BT$, the latter being the typical energy that can be removed from the sample in a cooling cycle. In this way, at a typical detuning of $k_BT/ h$ in frequency units to the red of the resonance, efficient excitation can be obtained. Given the effect of multiple particle collisions in this ultrahigh pressure regime, we expect that the cooling efficience will be stronger dependent on the buffer gas pressure than estimated from a purely binary collisions picture.

\section{Conclusions}
We have described an experiment demonstrating collisional redistribution cooling of an ultradense atomic gas mixture, rubidium atoms and 230\,bar argon buffer gas. In this proof of principle experiment a relative cooling by 66\,K was achieved, limited by the thermal conductivity of the argon gas. The cooled gas has a density exceeding 10$^{21}$ atoms/cm$^3$, corresponding to a laser cooling of macroscopic gas samples, and the cooling power is 87\,mW.

For the future, it will be interesting to explore the cooling limits of collisional redistribution laser cooling. The presently obtained temperature drop is fully consistent with being determined by the thermal conductivity of the buffer gas material. A simple estimate of the temperature drop in the beam focus, as limited by thermal conductivity, in a cylindrical geometry yields $\Delta T=\frac{P_{\textit{cool}}}{2\pi\kappa L}\cdot\ln{(r_e/r_i)}$, where $L \simeq l_{\textit{abs}}$ denotes the length of the cooling region, $r_i (\simeq \omega)$ and $r_e$ inner and outer radii of integration, and $\kappa$ the thermal conductivity of the gas. For the sake of simplicity, one may set $r_e\simeq l_{\textit{abs}}$, which gives a simple scaling formula for the expected cooling limt, as expected from thermal conductivity (see Eq.\,7 for a more exact description). We find that the expected minimum temperature can e.g. be improved by a decrease of the absorption length $l_{\textit{abs}}$ , as can be achieved by an enhanced rubidium vapor pressure. Also, a tighter beam focussing is required to maintain the limit $l_{\textit{abs}}\gg \omega$ of a cylindrical geometry, while radii variations enter only logarithmically into the expected minimum temperature. By use of a shorter absorption length (and a tighter beam focus), the available cooling power is deposited in a smaller volume, with correspondingly lower outer area through which the heat transport can proceed. We are presently investigating the use of a commercial cell, as described in chapter 3, which should be capable of operating at higher temperatures to increase the rubidium vapor pressure  (with respect to the results obtained in the home-made pressure cell, with which most of the here reported measurements have been taken); first encouraging results for the case of potassium vapor buffer gas mixtures in our experiment are presented in \cite{sass}. We note that an alternative option would be the use of other buffer gases, e.g. krypton or xenon, which have lower thermal conductivity than the presently used argon gas.

Given the high value of the saturation intensity in the pressure broadened ensemble, the cooling limit can also be extended by increasing the cooling laser power. The most straightforward approach here seems the use of high power multimode diode lasers, which are available up to kW power levels. While we have so far discussed cooling inside the gas, one may also investigate cooling of a thermally isolated all-sapphire pressure cell, which may allow for novel optical chillers \cite{Sheik-Bahae2}.

An interesting perspective of redistribution laser cooling is to investigate supercooling beyond the homogeneous nucleation temperature and related non-equilibrium phenomena \cite{kraska, fladerer}.
With the present results we have shown the ability to manipulate via laser radiation the local temperature in a gas sample of 230\,bar argon partial pressure in the temperature range between 570\,K-630\,K, which is well beyond the critical point of argon (151\,K, 48.7\,bar). Though recent experiments have demonstrated that even beyond the critical point gas and liquid phases can be distinguishable when crossing the Widom line \cite{Simeoni}, it seems feasible to extend the operating range of the here investigated cooling technique to below the critical pressure of the buffer gas. The noble gas with the 'nearest' critical point would be xenon ($T_c$=289.8\,K, $p_c$=58.4\,bar). While it might be preferable to work with higher buffer gas pressures than 58.4\,bar to achieve effective cooling, the use of a binary mixture of argon and xenon buffer gas could offer both features, a high pressure broadening to support the absorption of the cooling laser beam and an adjustable partial pressure of xenon. We are aware that in the binary noble gas mixture the phase boundaries are shifted. A further appealing feature of the presented cooling scheme is its speed. The characteristic time to establish a given temperature profile in the gas is $t_c=\frac{\omega^2}{4D}$ \cite{whinnery}. While the thermal diffusivity is mostly determined by the required gas pressure, the radius of the cooling beam $\omega$ can readily be decreased below the value of 1\,mm used in the present work. The resulting simultaneous decrease of the cooled volume and the chacteristic timescale could be an promising way to investigate the still somewhat elusive critical phenomena in the deeply supercooled regime, as the glass transition, which, as generic critical phenomenon, benefits experimentally from both small volumes and short timescales \cite{Kivelson, cavagna}.

\section*{Acknowledgements}
M. W. thanks Fr$\acute{\text{e}}$d$\acute{\text{e}}$ric Caupin for valuable discussions concerning nonequilibrium thermodynamics. Financial support from the Deutsche Forschungsgemeinschaft within the focused research unit FOR557 is acknowledged.


\end{document}